\documentclass[prd,aps,twocolumn,preprintnumbers, showpacs, nofootinbib,superscriptaddress,notitlepage]{revtex4-1}
\usepackage{mathrsfs}
\usepackage{amsfonts}
\usepackage{amsmath}
\usepackage{slashed}
\usepackage{array}
\usepackage{verbatim}
\usepackage{epsfig}
\usepackage{graphicx}
\usepackage{color}

\newcommand{\beq}{\begin{eqnarray}}
\newcommand{\eeq}{\end{eqnarray}}

\def \non {\nonumber}

\begin{document}

\title{Renormalization in Large Momentum Effective Theory of Parton Physics}

\author{Xiangdong Ji}
\affiliation{Tsung-Dao Lee Institute, and College of Physics and Astronomy, Shanghai Jiao Tong University, Shanghai, 200240, P. R. China}
\affiliation{Maryland Center for Fundamental Physics, Department of Physics, University of Maryland, College Park, Maryland 20742, USA}

\author{Jian-Hui Zhang}
\affiliation{Institut f\"{u}r Theoretische Physik, Universit\"{a}t Regensburg, D-93040 Regensburg, Germany}

\author{Yong Zhao}
\affiliation{Center for Theoretical Physics, Massachusetts Institute of Technology, Cambridge, MA 02139, USA}

\preprint{MIT-CTP/4916}

\begin{abstract}

In the large-momentum effective field theory approach to parton physics,
the matrix elements of non-local operators of quark and gluon fields, linked by straight
Wilson lines in 
a spatial direction, are calculated in lattice quantum chromodynamics as a function of hadron momentum.
Using the heavy-quark effective theory formalism, we show a multiplicative renormalization
of these operators at all orders in perturbation theory, both in dimensional and lattice
regularizations. The result provides a theoretical basis for
extracting parton properties through properly renormalized observables
in Monte Carlo simulations.

\end{abstract}

\maketitle

\section{introduction}

One of the most important goals of quantum chromodynamics (QCD) is to understand the hadron structure from its fundamental degrees of freedom --- quarks and gluons.
A powerful tool of obtaining such results is lattice QCD, which has been used to study the static properties of the hadron, such as mass, charge radius etc., to good accuracy~(see e.g.~\cite{Edwards:2012fx,Aoki:2016frl}). However, in general it cannot be used to directly access quantities with real-time dependence. An example is the parton distribution functions (PDFs), which are defined as the quark and gluon matrix elements of light-cone correlations and describe the momentum distribution of quarks and gluons inside the hadron. They play a crucial role in understanding the experimental data at high energy hadron colliders such as the LHC. Lattice QCD can only be used to indirectly extract the information of the PDFs by calculating their moments~\cite{Detmold:2001jb,Dolgov:2002zm}, which is limited by technical difficulties. 

In the past few years, a new approach has been proposed to study parton physics from lattice QCD, which is now known as the large-momentum effective theory (LaMET)~\cite{Ji:2013dva,Ji:2014gla}. Here one starts from a time-independent quasi-PDF, which is a matrix element of non-local quark and gluon operators that can be directly calculated on the lattice, and then uses it to extract the light-cone PDF through a perturbative factorization up to power corrections suppressed by the nucleon momentum. The factorization in Refs.~\cite{Ji:2013dva,Xiong:2013bka} was given for the bare quark quasi-PDF, where all fields and couplings entering the quasi-distribution are bare ones.
The ultraviolet (UV) physics of the quasi-PDF is all factorized perturbatively into the matching coefficients. However, lattice perturbation theory is known to converge slowly, and non-perturbative renormalization is often necessary to define a continuum limit where the matching to
parton physics is under better control. Thus to make LaMET more practical, a complete understanding of UV divergences of the relevant non-local operators
is essential.

Renormalization of non-local operators in QCD has not been well studied in the literature. A distinct feature of the LaMET non-local operators is the appearance of Wilson lines connecting the parton fields at different spacetime points to ensure gauge invariance.
The renormalization of an open Wilson line as well as a closed Wilson loop has been studied decades ago~\cite{Dotsenko:1979wb,Craigie:1980qs,Dorn:1986dt}, where it was shown that the Wilson line with a smooth simple contour can be renormalized, and in particular, the power divergence associated with the Wilson line can be absorbed into an exponential mass renormalization. In an auxiliary field formalism~\cite{Dorn:1986dt}, it was also shown that the Wilson line can be replaced by a Green's function of the auxiliary field and studied as such. On the other hand, the renormalization of non-local operators such as the quasi-PDF has extra complications due to the parton fields attached to the endpoints.  
We have shown previously that the quasi-PDF is multiplicatively renormalizable in dimensional regularization up to two-loop order~\cite{Ji:2015jwa}.
The full renormalization properties of the quasi-PDF has been conjectured in recent studies~\cite{Ishikawa:2016znu,Chen:2016fxx,Constantinou:2017sej,Alexandrou:2017huk,Chen:2017mzz,Stewart:2017ss} (for an earlier work on the renormalization of transverse-momentum-dependent distributions with a similar conjecture, see~\cite{Musch:2010ka}),
but not proven.

In this paper, we perform a systematic study of the renormalization of gauge-invariant non-local operators for the quasi-PDF. We first present a general framework by introducing an auxiliary ``heavy quark" field into the QCD Lagrangian, where the Wilson line can be replaced by a product of the auxiliary fields. In analogy to heavy-quark effective theory (HQET), we argue that this theory is renormalizable to all orders in perturbation theory. By taking the unpolarized quark quasi-PDF as an example, we then show that its renormalization reduces to that of two heavy-light quark currents, which can also be done to all orders in perturbation theory. This removes
the obstacle to define a continuous quasi-PDF through lattice simulations. The same conclusion is not limited to the quark quasi-PDF, but can also be generalized to the gluon ones, as well as
other non-local correlators, such as the Euclidean observables defined in the LaMET framework to extract the generalized parton distributions,
transverse-momentum-dependent distributions, hadron distribution
amplitudes, etc. 

\section{Effective QCD with Auxiliary ``Heavy-Quark" Field}

To be concrete, consider the following non-local operator
\begin{equation}\label{Eq:nonlocalop}
 O(x,y) = \overline{\psi}(x) \Gamma L(x,y)\psi(y),
\end{equation}
where $\psi, \bar\psi$ denote the bare quark fields and $\Gamma$ a Dirac structure. $L(x, y)$ is a path-ordered gauge link from $y$ to $x$,
\begin{equation}
             L(x, y) = {\cal P}\exp\left(-ig\int^1_0 d\lambda\,\frac{dz^\mu}{d\lambda}  A_\mu(z(\lambda))\right)
\end{equation}
carrying a $(3,\bar 3)$ representation in color SU(3) group, where the path is parametrized such that $z(0)=y$ and $z(1)=x$. The gauge link ensures gauge invariance of the operator $O(x,y)$. 
For a straight-line gauge link which is of our main interest in the present paper, we can choose $z^\mu(\lambda)=y^\mu+\eta\lambda\, n^\mu$ with $n^\mu$ a unit vector specifying the direction and $\eta$ the length of the gauge link.

To study the renormalization property of the above operator, we introduce an ``heavy quark" auxiliary field (color source without spin degrees of freedom)
$Q$ with color 3,
and its conjugate ${\overline Q}$, with color ${\bar 3}$. 
We extend QCD to include this ``heavy quark" interaction with the gluon field,  and introduce the following Lagrangian
\beq\label{Eq:lag}
{\cal L}={\cal L_{\text{QCD}}}+\overline Q(x)i n\cdot D Q(x),
\eeq
where $D_\mu=\partial_\mu+igt^a A^a_\mu$ is the covariant derivative in fundamental representation. For a real heavy quark or timelike Wilson line, $n^\mu$ is a timelike vector $n^\mu=(1,0,0,0)$, whereas for a spacelike Wilson line $n^\mu$ can be chosen as $n^\mu=(0,0,0,1)$. We will focus on the latter case in the following, although the discussion may go through for any $n$.
So long as there is a non-zero time component $n^0$, the ``heavy quark" is dynamical, but follows a designated world line.

In the above theory, we can replace the bilocal operator $O(x,y)$ by a new operator,
\begin{equation}\label{Eq:repl}
             O(x, y) = \overline{\psi}(x)\Gamma Q(x) \overline{Q}(y) \psi(y).
\end{equation}
This can be seen as following. After integrating out the ``heavy quark" field, we have
\beq
\int {\cal D}\overline Q{\cal D}Q\, Q(x)\overline Q(y) e^{i\int d^4x {\cal L}}=S_Q(x, y) e^{i\int d^4x {\cal L_{\text{QCD}}}}
\eeq
up to a determinant ${\rm det}(in\cdot D)$ which can be shown to be a constant and absorbed into the normalization of the generating functional~\cite{Mannel:1991mc}, where $S_Q(x, y)$ satisfies
\beq
n\cdot D\, S_Q(x, y)=\delta^{(4)}(x-y),
\eeq
with the solution in the case of $n^\mu=(0,0,0,1)$
\begin{align}
&S_Q(x,y)=\theta(x^z-y^z)\delta(x^0-y^0)\delta^{(2)}(\vec x_\perp-\vec y_\perp)L(x,y)\non\\
&=\theta(x^z-y^z)\delta(x^0-y^0)\delta^{(2)}(\vec x_\perp-\vec y_\perp)L(x^z,y^z),
\end{align}
under an appropriate boundary condition. In the following we will restrict to the case $x^z>y^z$ without loss of generality. The $\delta$-functions ensure that the time and transverse components of $x$ and $y$ are equal, and therefore generate a spacelike Wilson line along the longitudinal direction. The above result allows us to replace the Wilson line $L(x^z,y^z)$, which is the one appearing in the quasi-PDF, by the product of two auxiliary heavy quark fields $Q(x^z)\overline Q(y^z)$. The non-local operator in the quasi-PDF then becomes the product of two composite operators in Eq.~(\ref{Eq:repl}).

\section{Renormalization of Effective QCD with ``Heavy Quark" in Dimensional Regularization}

Let us consider renormalization of the above effective theory with a ``heavy quark".
We will show that such a theory can be renormalized perturbatively to all orders in perturbation theory,
first in dimensional regularization where power divergences do not exist. The basic argument
is the same as the all-order proof of the renormalization for HQET, first
presented in Ref.~\cite{Bagan:1993zv}.

The standard proof of the renormalization of QCD chooses the covariant gauge $\partial_\mu A^\mu = 0$, implemented
in the functional integrals with a gauge parameter $\xi$. To recover a local theory after gauge
fixing, one has to introduce the color-octet ghost fields $c^a$ and $\bar c^a$. The resulting theory
has a residual global symmetry called BRST symmetry. The Ward-Takahashi identities
from this symmetry are a key ingredient for showing the theory is renormalizable to all-orders
in perturbation theory in the following sense: After absorbing all the infinities into
wave function renormalization factors of quark ($Z_1$), gluon $(Z_3$), and ghost fields $(Z_c)$, and the multiplicative
renormalization of quark masses $(Z_m)$, gauge coupling ($Z_g$) and gauge fixing parameter ($Z_\xi$),
Green's functions of the renormalized elementary fields are UV finite.

In HQET where the Lagrangian has a similar form as Eq.~(\ref{Eq:lag}) except that the vector $n^\mu$ is timelike instead of spacelike, it has been shown that to leading order in the heavy quark mass the Green's functions can be renormalized to all-orders in perturbation theory~\cite{Bagan:1993zv}, where a BRST invariance of the HQET Lagrangian was used. In HQET, heavy quarks do not couple to heavy antiquarks, hence no heavy quark loops can occur. This implies that Green's functions without external heavy quark legs can be renormalized in the same way as in QCD. Only those involving external heavy quark legs need to be considered separately, which yield
a new wave function renormalization constant ($Z_Q$). These Green's functions include the heavy-quark two-point function, the heavy-quark-gluon vertex and the heavy-quark-ghost vertex with one insertion of composite operators coming from the BRST transformation of the heavy quark field. All of them can be shown to be UV finite with the help of Ward-Takahashi identities under the given finite number of renormalization factors in the HQET Lagrangian~\cite{Bagan:1993zv}.

Note that the above arguments are valid independent of whether the heavy quark field in the HQET Lagrangian is defined by a timelike or spacelike vector.
Therefore, the statement about renormalizability of the HQET in Ref.~\cite{Bagan:1993zv} shall also be
true for our auxiliary heavy quark Lagrangian with $n^\mu=(0,0,0,1)$ in Eq.~(\ref{Eq:lag}). In the case of light-cone vector, however,
new divergences appear due to light-cone singularities.

Local composite operators in this effective theory can be renormalized using the standard approach. In particular, the heavy-light composite operators, such as
\begin{equation}
             \overline{j}(x) = \overline{\psi}(x)\Gamma Q(x),
\end{equation}
acquires a divergent factor $Z_j$ (Note that $\overline{j}(x)$
now carries a spinor index). In analogy to the Green's functions, one can show that the overall UV divergence of the insertion of $\bar{j}(x)$ into Green's functions is subtracted by $Z_j$ to all orders of perturbation theory. Therefore, we can write $\overline{j}= Z_j \overline{j}_R$ where $\overline{j}_R$
is a renormalized operator.
 The calculation of its anomalous dimension
has been worked out in the HQET~\cite{Politzer:1988wp,Ji:1991pr}. 
For the auxiliary heavy quark field, actually one can explicitly verify that at one-loop in dimensional regularization (with $D=4-2\epsilon$)
\begin{equation}\label{eq:hl}
   Z_j = 1+ \frac{\alpha_s}{2\pi\epsilon} \ ,
\end{equation}
which is the same as in HQET.

\section{Renormalization of Non-Local Operators in Dimensional Regularization}

Now let us consider the renormalization of the non-local operator, such as in Eq.~(\ref{Eq:nonlocalop}), which appears
in the LaMET approach to parton physics. In the effective theory, the non-local operator
becomes a product of local composite operators,
\begin{equation}
 O(z_2, z_1) = \overline{j}(z_2) j(z_1),
\end{equation}
after we replace the non-local Wilson line by the product of two auxiliary ``heavy quark" fields.
 This operator can be multiplicatively renormalized by
 \begin{equation}
        O(z_2,z_1)= Z_{\bar j}Z_{j} O_R(z_2,z_1)
\end{equation}
to all orders in perturbation theory with $Z_{j}=Z_q^{1/2} Z_Q^{1/2} Z_{V_j}$, where $Z_q, Z_Q, Z_{V_j}$ are the renormalization constant for the light-quark, heavy-quark and vertex, respectively.

Thus the renormalized operator becomes $O_R=Z_{\bar j}^{-1}Z_{j}^{-1}O(z_2,z_1)$. After integration over the ``heavy quark" field, we have
\begin{equation}\label{Eq:Orenorm}
        O_R = Z_{\bar j}^{-1}Z_{j}^{-1} \overline{\psi}(z_2) \Gamma L(z_2,z_1)\psi(z_1),
\end{equation}
where all fields on the r.h.s. are bare fields. In Ref.~\cite{Ji:2015jwa}, it has been shown that the quasi-PDF operator requires the renormalization of virtual diagrams only, which is equivalent to the result in Eq.~(\ref{Eq:Orenorm}) that we only need renormalization of the heavy-light quark currents for the operator $O$. In particular, the one-loop renormalization factor for the quasi-PDF
\beq\label{Eq:1loopZfac}
Z_O=1+\frac{\alpha_s }{\pi\epsilon}
\eeq
is consistent with the anomalous dimension of the heavy-light quark current in Eq.~(\ref{eq:hl}). The same is true at the two-loop order~\cite{Ji:2015jwa}.

\section{Renormalization of Non-Local Operators in Lattice Regularization}

The HQET Lagrangian in Eq.~(\ref{Eq:lag}) takes the infinite heavy quark mass limit,
therefore does not contain any mass term. The mass correction to the heavy quark can only receive power divergent contributions~\cite{Maiani:1991az}, which are absent in dimensional regularization.

However, in UV cut-off regularizations such as the lattice regularization,
when going beyond leading-order perturbation theory, the self-energy of the heavy quark introduces a linear divergence which
has to be absorbed into an effective mass counterterm,
\begin{equation}
    \delta {\cal L}_m =  -\delta m\overline{Q} Q
\end{equation}
with $\delta m$ of ${\cal O}(1/a)$ with $a$ the lattice spacing~\cite{Maiani:1991az}. Although lattice regularization breaks Lorentz symmetry and introduces operator mixing, it is not a concern for $O(z_2,z_1)$ because there is no lower-dimension operator that can mix with it in lattice QCD.
In the HQET framework, the appearance of the linear divergence is easy to understand. Given that the heavy quark is infinitely heavy, it behaves like a static color source, whose energy will have a Coulomb-like form $1/r$, and therefore will diverge linearly if the source is a pointlike particle.
In our auxiliary heavy quark Lagrangian, the physical picture as a static color source is lost, whereas the removal of linear divergence can be done essentially in the same way. Since the linear divergence appears only in the self energy of the auxiliary heavy quark, it can be absorbed into the renormalization of the heavy quark mass as in Ref.~\cite{Maiani:1991az}. 

Thus, the total heavy-quark Lagrangian now becomes
\begin{equation}\label{Eq:spaceliken}
    {\cal L}_Q = \bar Q (i n\cdot D-\delta m)Q.
\end{equation}
Note that the BRST invariance of the Lagrangian requires a dependence of $\delta m$ on the signature of $n$~\cite{Dorn:1986dt}, which yields a vanishing $\delta m$ for a lightlike $n^\mu$. In the case of a spacelike $n^\mu$, $\delta m$ in Eq.~(\ref{Eq:spaceliken}) is imaginary, we can therefore write $\delta m=i\delta {\bar m}$. After integrating over the heavy fields,  the renormalized operator becomes,
\begin{equation}\label{Eq:cutoffrenorm}
        O_R = Z_{\bar j}^{-1}Z_{j}^{-1}e^{\delta \bar m|z_2-z_1|}\overline{\psi}(z_2) \Gamma L(z_2,z_1)\psi(z_1).
\end{equation}
Thus there is a mass counterterm in the exponent which has a $z$-dependence.
After the mass renormalization, there are at most logarithmic divergences which will be canceled by counterterms from the Lagrangian or from the renormalization of the heavy-light composite operator. The above renormaliztion form has been proposed in a number of papers before~\cite{Musch:2010ka,Ishikawa:2016znu,Chen:2016fxx,Constantinou:2017sej,Alexandrou:2017huk,Chen:2017mzz},
but not proven.

It is worthwhile to point out that the explicit one-loop calculation~\cite{Chen:2016fxx} in lattice regularization indeed shows that the exponential mass renormalization in the form of Eq.~(\ref{Eq:cutoffrenorm}) cancels the linear divergence in the quasi-PDF in Ref.~\cite{Xiong:2013bka} after a Fourier transform to momentum space, where
\beq
\delta \bar m=\frac{2\pi\alpha_s }{3a}.
\eeq

The renormalization presented above works the same when $z_2<z_1$.

\section{conclusion}
In this paper, we have shown the non-local operator involving a spacelike Wilson line calculated in LaMET can be renormalized multiplicatively to all-orders in perturbation theory. Apart from the mass counterterm which yields
a $z$-dependent exponential, all the other renormalization constants are independent of $z$ and related to the renormalization
of the heavy-light current in the HQET. This provides an important theoretical basis for defining the
continuum limit of the quasi-PDFs in lattice QCD, which is necessary for precision calculations of parton physics
using the LaMET approach.

Note: When this article is being finalized, we learnt that a similar consideration has been made by the DESY group~\cite{greentalk}.

\section*{Acknowledgement}
This work was partially supported by the U.S. Department of Energy Office of Science, Office of Nuclear Physics under Award Number DE-FG02-93ER-40762 and DE-SC0011090, a grant from the National Science Foundation of China (No. 11405104) and the SFB/TRR-55 grant ``Hadron Physics from Lattice QCD". The work of XJ and YZ was also supported in part by the U.S. Department of Energy, Office of Science, Office of Nuclear Physics, within the framework of the TMD Topical Collaboration.


\begin{thebibliography}{99}

\bibitem{Edwards:2012fx}
  R.~G.~Edwards {\it et al.} [Hadron Spectrum Collaboration],
  Phys.\ Rev.\ D {\bf 87}, no. 5, 054506 (2013)
  doi:10.1103/PhysRevD.87.054506
  [arXiv:1212.5236 [hep-ph]].

\bibitem{Aoki:2016frl}
  S.~Aoki {\it et al.},
  Eur.\ Phys.\ J.\ C {\bf 77}, no. 2, 112 (2017)
  doi:10.1140/epjc/s10052-016-4509-7
  [arXiv:1607.00299 [hep-lat]].


	\bibitem{Detmold:2001jb}
	W.~Detmold, W.~Melnitchouk, J.~W.~Negele, D.~B.~Renner and A.~W.~Thomas,
	Phys.\ Rev.\ Lett.\  {\bf 87}, 172001 (2001)
	doi:10.1103/PhysRevLett.87.172001
	[hep-lat/0103006].
	
	
	\bibitem{Dolgov:2002zm}
	D.~Dolgov {\it et al.} [LHPC and TXL Collaborations],
	Phys.\ Rev.\ D {\bf 66}, 034506 (2002)
	doi:10.1103/PhysRevD.66.034506
	[hep-lat/0201021].
	

	\bibitem{Ji:2013dva}
	X.~Ji,
	Phys.\ Rev.\ Lett.\  {\bf 110}, 262002 (2013)
	doi:10.1103/PhysRevLett.110.262002
	[arXiv:1305.1539 [hep-ph]].
	
	
	\bibitem{Ji:2014gla}
	X.~Ji,
	Sci.\ China Phys.\ Mech.\ Astron.\  {\bf 57}, 1407 (2014)
	doi:10.1007/s11433-014-5492-3
	[arXiv:1404.6680 [hep-ph]].
	
	
	\bibitem{Xiong:2013bka}
	X.~Xiong, X.~Ji, J.~H.~Zhang and Y.~Zhao,
	Phys.\ Rev.\ D {\bf 90}, no. 1, 014051 (2014)
	doi:10.1103/PhysRevD.90.014051
	[arXiv:1310.7471 [hep-ph]].
	
\bibitem{Dotsenko:1979wb}
  V.~S.~Dotsenko and S.~N.~Vergeles,
  Nucl.\ Phys.\ B {\bf 169}, 527 (1980).
  doi:10.1016/0550-3213(80)90103-0

\bibitem{Craigie:1980qs}
  N.~S.~Craigie and H.~Dorn,
  Nucl.\ Phys.\ B {\bf 185}, 204 (1981).
  doi:10.1016/0550-3213(81)90372-2

\bibitem{Dorn:1986dt}
  H.~Dorn,
  Fortsch.\ Phys.\  {\bf 34}, 11 (1986).
  doi:10.1002/prop.19860340104

\bibitem{Ji:2015jwa}
  X.~Ji and J.~H.~Zhang,
  Phys.\ Rev.\ D {\bf 92}, 034006 (2015)
  doi:10.1103/PhysRevD.92.034006
  [arXiv:1505.07699 [hep-ph]].

\bibitem{Ishikawa:2016znu}
  T.~Ishikawa, Y.~Q.~Ma, J.~W.~Qiu and S.~Yoshida,
  arXiv:1609.02018 [hep-lat].

\bibitem{Chen:2016fxx}
  J.~W.~Chen, X.~Ji and J.~H.~Zhang,
  Nucl.\ Phys.\ B {\bf 915}, 1 (2017)
  doi:10.1016/j.nuclphysb.2016.12.004
  [arXiv:1609.08102 [hep-ph]].

\bibitem{Constantinou:2017sej}
  M.~Constantinou and H.~Panagopoulos,
  arXiv:1705.11193 [hep-lat].

\bibitem{Alexandrou:2017huk}
  C.~Alexandrou, K.~Cichy, M.~Constantinou, K.~Hadjiyiannakou, K.~Jansen, H.~Panagopoulos and F.~Steffens,
  arXiv:1706.00265 [hep-lat].


\bibitem{Chen:2017mzz}
J.~W.~Chen, T.~Ishikawa, L.~Jin, H.~W.~Lin, Y.~B.~Yang, J.~H.~Zhang and Y.~Zhao,
arXiv:1706.01295 [hep-lat].


	\bibitem{Stewart:2017ss}
	I.~Stewart and Y.~Zhao,
	to be published soon.


\bibitem{Musch:2010ka} 
  B.~U.~Musch, P.~Hagler, J.~W.~Negele and A.~Schafer,
  Phys.\ Rev.\ D {\bf 83}, 094507 (2011)
  doi:10.1103/PhysRevD.83.094507
  [arXiv:1011.1213 [hep-lat]].



\bibitem{Mannel:1991mc}
  T.~Mannel, W.~Roberts and Z.~Ryzak,
  Nucl.\ Phys.\ B {\bf 368}, 204 (1992).
  doi:10.1016/0550-3213(92)90204-O


\bibitem{Bagan:1993zv}
  E.~Bagan and P.~Gosdzinsky,
  Phys.\ Lett.\ B {\bf 320}, 123 (1994)
  doi:10.1016/0370-2693(94)90834-6
  [hep-ph/9305297].

\bibitem{Politzer:1988wp}
  H.~D.~Politzer and M.~B.~Wise,
  Phys.\ Lett.\ B {\bf 206}, 681 (1988).
  doi:10.1016/0370-2693(88)90718-6



\bibitem{Ji:1991pr}
  X.~D.~Ji and M.~J.~Musolf,
  Phys.\ Lett.\ B {\bf 257}, 409 (1991).
  doi:10.1016/0370-2693(91)91916-J


\bibitem{Maiani:1991az}
  L.~Maiani, G.~Martinelli and C.~T.~Sachrajda,
  Nucl.\ Phys.\ B {\bf 368}, 281 (1992).
  doi:10.1016/0550-3213(92)90528-J


\bibitem{greentalk}
  J.~Green,
  {\href{https://makondo.ugr.es/event/0/session/95/contribution/332/material/slides/0.pdf}{\it Auxiliary field approach to extended operators for quasi-PDFs}},
  (Talk given at the 35th International Symposium on Lattice Field Theory, Granada, Spain, 2017).


\end{thebibliography}
\end{document}